\documentclass[aps,twocolumn,floatfix,superscriptaddress,letterpaper,prb,nofootinbib,showpacs]{revtex4}
\usepackage{graphicx}
\usepackage{amssymb}
\usepackage{verbatim}
\usepackage{amsfonts}

\begin{document}

\title{Cryogenic ion trapping systems with surface-electrode traps}

\author{P. B. Antohi\footnote{Electronic mail: {apaul@mit.edu}}}
\affiliation{Center for Ultracold Atoms, Department of Physics, Massachusetts Institute of Technology, Cambridge, MA
02139, USA}

\author{D. Schuster}
\affiliation{Department of Applied Physics, Yale University, New Haven, CT 06511, USA}

\author{G. M. Akselrod}
\affiliation{Center for Ultracold Atoms, Department of Physics, Massachusetts Institute of Technology, Cambridge, MA
02139, USA}

\author{J. Labaziewicz}
\affiliation{Center for Ultracold Atoms, Department of Physics, Massachusetts Institute of Technology, Cambridge, MA
02139, USA}

\author{Y. Ge}
\affiliation{Center for Ultracold Atoms, Department of Physics, Massachusetts Institute of Technology, Cambridge, MA
02139, USA}

\author{Z. Lin}
\affiliation{Center for Ultracold Atoms, Department of Physics, Massachusetts Institute of Technology, Cambridge, MA
02139, USA}

\author{W. S. Bakr}
\affiliation{Center for Ultracold Atoms, Department of Physics, Harvard University, Cambridge, MA 02138, USA}

\author{I. L. Chuang}
\affiliation{Center for Ultracold Atoms, Department of Physics, Massachusetts Institute of Technology, Cambridge, MA
02139, USA}

\date{\today}

\begin{abstract}
We present two simple cryogenic RF ion trap systems
in which cryogenic temperatures and ultra high vacuum pressures can 
be reached in as little as 12 hours. The ion traps are operated either 
in a liquid helium bath cryostat or in a low vibration closed cycle 
cryostat. The fast turn around time and availability of buffer gas 
cooling made the systems ideal for testing surface-electrode ion traps.
The vibration amplitude of the closed cycled cryostat was found to be
below 106 nm. We evaluated the systems by loading surface-electrode
ion traps with $^{88}$Sr$^+$ ions using laser ablation, which is 
compatible with the cryogenic environment. Using Doppler cooling we 
observed small ion crystals in which optically resolved ions have a 
trapped lifetime over 2500 minutes.
\end{abstract}


\maketitle

\tableofcontents

\section{Introduction}

In this paper we detail the construction and characterize the performance of two cryogenic ion traps systems, designed for trapping $^{88}$Sr$^+$ ions. Ion traps operated at liquid helium temperature offer several advantages and pose some challenges compared with room temperature ultra high vacuum (UHV) systems. The cryogenic environment produces high vacuum and subsequently reduces the collision rate with the background gas.\cite{Gab} Moreover the low temperatures reduce the electrical (Johnson) noise and the anomalous heating of trapped ions.\cite{Des,Die,Des1,Roo,Tur,Jarek,Dav1} Cryogenic systems also present some technical advantages compared with room temperature systems. The time necessary to reach UHV pressures can be 12 hours or less. Since the cryogenic temperature suppresses outgassing, the range of materials that can be used for the trap construction is increased, as they do not have to be UHV compatible. However, some of the methods used in room temperature ion trap systems are not directly transferable to a cryogenic system, since careful attention must be paid in how the cryogenic system handles the thermal loads.
One important problem with the standard high-temperature ion sources is that they are not compatible with the cryogenic environment due to the high heat load. We found that a compatible ion source can be obtained through a laser ablation method, which was demonstrated in earlier experiments.\cite{Kwo,Gil,Dec,Has,KITA} Though care must be taken to provide appropriate thermal anchoring and isolation, it is possible to use all conventional ion sources, including photoionization and electron impact ionization of atoms produced by an oven. In this paper we focus on laser ablation because it is simple and able to load a variety of species.
Another problem is that the RF dissipation can locally heat the trap electrodes, which in the case of a cryogenic system requires a very good thermal contact to the cryostat baseplate in order dissipate the generated heat. To address this problem we used optical lithography techniques to fabricated a surface-electrode ion trap, since the planar aspect of this trap simplifies the anchoring to the helium baseplate. In addition, surface-electrode traps are easily integrable in multiplexed ion traps as a proposed scheme for the realization of a large-scale quantum computer.\cite{Kie,Sch,Chi,Pea,Sei}

Cryogenic ion traps allow for exploring new physics, such as the quantum interaction between superconductors and ions.\cite{Tia} Proposals for a quantum information processor based on trapped ions require the ions to be cooled to the motional ground state, and decoherence rates in current ion traps may limit the fidelity of quantum operations.\cite{Cir,Win,Jam} Some of these decoherence sources can be minimized by the cryogenic environment. One source is the motional heating caused by electrical noise in the trap electrodes. Studies of the dependence of the ion heating rates on the trap dimensions show that the main component of the heating is anomalous heating which dominates the Johnson noise.\cite{Des,Die,Des1,Roo,Tur} Previous experiments with the trap electrodes cooled to 150 K showed a reduction of the anomalous heating rate at least one order of magnitude compared with electrodes kept at room temperature.\cite{Des} Recent experiments with the ion trap electrodes cooled to 6 K showed an anomalous heating rate reduction of between seven and eight orders of magnitude.\cite{Jarek} Another cause for decoherence is collisions with the background gas molecules, which alter the internal quantum state of a trapped ion.\cite{Win} Elastic collisions increase the kinetic energy the trapped ion that may eject the ion from the ion trap. Inelastic collisions such as chemical reactions and charge exchange can produce a different ion species or neutralize the trapped ion.

Several other cryogenic ion traps were successfully operated in previous experiments in order to take advantage of the very high vacuum and the superconducting properties of some materials offered by the cryogenic environments. A cryogenic linear RF ion trap for trapping
$^{199}$Hg$^+$ was built by the NIST ion storage group \cite{Poi} and used as a frequency standard. A similar system was
employed by Okada {\it et al.} for studying the Bohr-Weisskopf effect in unstable Be$^+$ isotopes.\cite{Oka} Cryogenic
Penning ion traps were used by Gabrielse {\it et al.} to trap antiprotons and carry out accurate measurements of the
antiproton inertial mass.\cite{Gab} Also, cryogenic systems were built for trapping neutral atoms. Willems and Libbrecht
\cite{Wil} trapped Cs atoms in cryogenic magneto-optical and magnetic traps, and more recently Nirrengarten {\it et al.}
\cite{Nir} used a superconducting atom chip to trap Rb atoms.

This work is a continuation of D. Berkeland's practice of building simple ion trap systems.\cite{Berk} We extend it to include ion traps operated in cryogenic environments. We build our cryogenic ion trap systems around two commercially available cryostats: a liquid helium bath cryostat and a low vibration closed cycled cryostat.  Depending on the  experiment, each cryostat presents some benefits over the other. For example, the bath cryostat is naturally vibration free while the closed cycled cryostat does not require a constant supply of cryogens or long term supervision.
In order to reduce the mechanical complexity of the setups we kept the total number of the components to a minimum. For this reason, for instance, we opted not to use a hermetically sealed pill-box, which is usually employed in cryogenic systems to improve the vacuum of the working chamber.\cite{Poi} We find that the trapped ions lifetime is not adversely affected by the choice of using a single vacuum space defined by the cryostat outer walls.

The rest of the paper is organized in two main sections. In the following section (Section~\ref{sec:csetup}) we make a detailed description of the components involved in the construction of the two cryogenic systems, along with their principles of operation. In the last section (Section~\ref{sec:eval}) we finish with a performance evaluation of these two cryogenic systems built around the bath cryostat and closed cycle cryostat.

\section{The cryostat setup}
\label {sec:csetup}

A good cryogenic ion trapping system must present some of the same
characteristics as a room temperature system such as high vacuum and
an easy method for ion loading. A low temperature system also brings
new benefits like the suppression of electrical noise and a faster
turnaround time. On the other hand, the cryogenic systems present few
challenges which are not present in the room temperature systems. The
cryogenic systems have a limited thermal load capacity, thus special
attention must be paid to reduce the heat loads from radiation,
conduction and internal heat sources. Specifically for the cryogenic
ion traps connection wires with low thermal conductivity are required,
and they have to be heat-sunk at each temperature stage of the
cryostat. In the case of the surface-electrode ion trap due to the
dielectric RF dissipation in the trap substrate, the trap must be
carefully thermally anchored to the 4.2 K substrate. Another problem
which must be addressed in the closed cycle cryostats due to their
principle of operation, is the damping of vibration.

\begin{figure}[htbp]
\includegraphics[width=2.9 in]{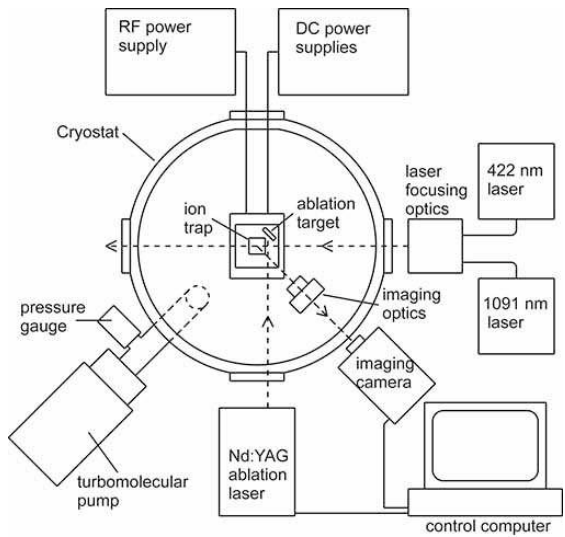}
\caption{Principal components used in the design of the cryostat systems. The fluorescence light is collected through the bottom viewport.} \label{Fig_1}
\end{figure}

In this section, we describe two cryostat setups we have constructed
and evaluated for $\sim $4 K operation of surface-electrode ion traps.
First, in Section~\ref{subsec:bath}, we describe a bath type cryostat system,
the workhorse for the majority of the experimental results presented
in Section~\ref{sec:eval}.  We have also constructed and evaluated a
closed cycle cryostat system, which has the advantage that it does not
necessitate a constant supply of cryogens. This cryostat design is described in Section~\ref{subsec:ccycle}. Both systems employ similar optical and control electronics subsystems, as described in the block diagram shown in Figure~\ref{Fig_1}.

\medskip

\subsection{The bath cryostat}
\label{subsec:bath}

\begin{figure}[htbp]
\includegraphics[width=2.9 in]{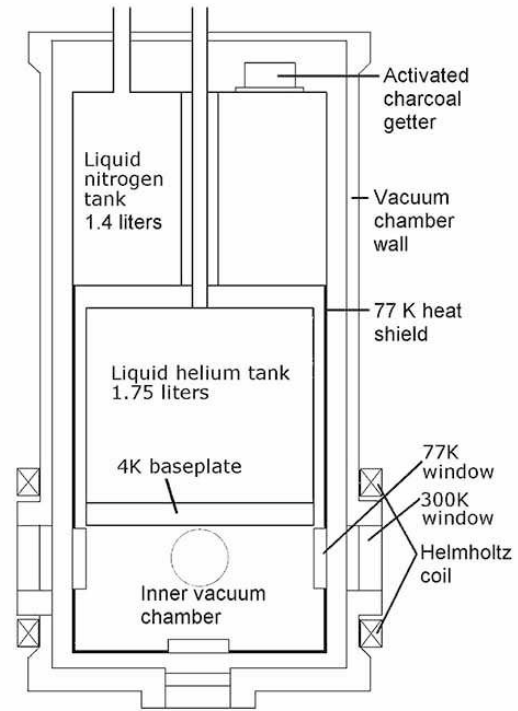}
\caption{Schematic diagram of the bath cryostat vertical cross section. The side windows are used for lasers access and the bottom viewport for fluorescent light collection. The ion trap with its components are not shown. (Not to scale)} \label{Fig_2}
\end{figure}

\begin{figure}[htbp]
\includegraphics[width=2.9 in]{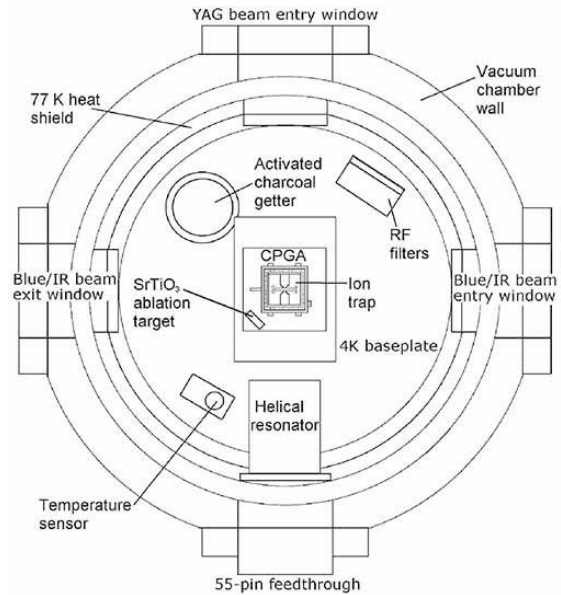}
\caption{Ion trap and the related components mounted on the 4.2 K baseplate of the bath cryostat. (Not to scale)} \label{Fig_3}
\end{figure}

The bath cryostat system is built from a commercial liquid helium bath
cryostat (QMC Instruments, Model TK 1813), depicted in
Figure~\ref{Fig_2}. The trap and related components are mounted on the
4.2 K baseplate of the cryostat (Figure~\ref{Fig_3}). The baseplate is heat sunk to the
outside of a 1.75 liter liquid helium reservoir that is surrounded by
a 77 K shield connected to a 1.4 liter liquid nitrogen tank. In this
setup the 77 K inner chamber is open to the main vacuum chamber.  The
temperature of the helium baseplate is monitored with a ruthenium
oxide sensor (LakeShore, RX-103A-AA). The sensor is mounted in small
block of OFHC copper, which is anchored to the baseplate. To enhance
thermal contact between the components mounted on the 77 K heat shield
or on the 4.2 K baseplate Apiezon N high vacuum grease is used.

\begin{figure}[htbp]
\includegraphics[width=2.9 in]{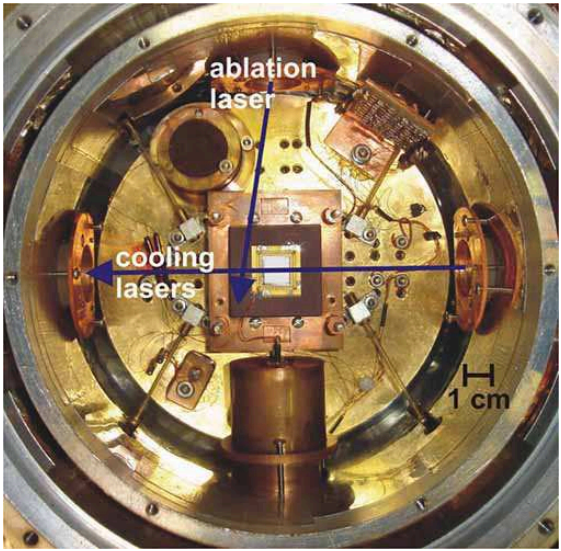}
\caption{View of the bath cryostat working chamber. The ablation target is not installed; the rest of visible components are described in Figure \ref{Fig_3}.} \label{Fig_4}
\end{figure}

\subsubsection{Ion trap and electrical components}

\begin{figure}[htbp]
\includegraphics[width=2.9 in]{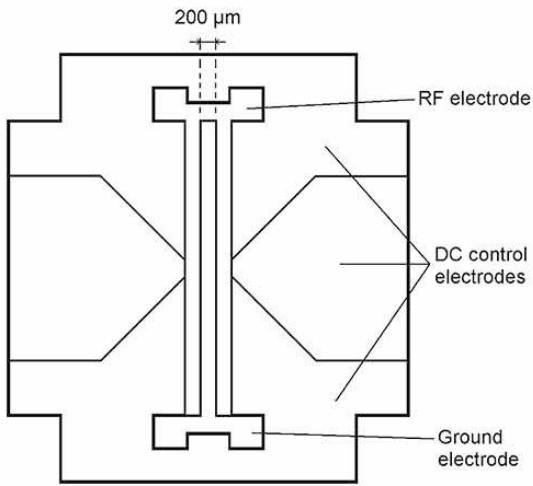}
\caption{Schematic diagram of the ion trap. The exposed substrate area is minimizated in order to reduce the accumulation of electrostatic charges. (To scale)} \label{Fig_5}
\end{figure}

For the bath cryostat system we constructed a surface-electrode ion trap, with five DC control electrodes and two RF electrode.\cite{Jarek} Compared with the traditional linear RF Paul trap with 3D electrode geometry,\cite{Berk} the surface-electrode ion trap has all the electrodes in a single plane. Also, the planar aspect of the trap offers the possibility to design complex electrode geometries.
We fabricate the traps by evaporating a 2 $\mu$m silver thin film on a sapphire crystal substrate, followed by a wet etch process in which the electrodes are developed. The width of the ground electrode is 200 $\mu$m with a spacing between RF and ground electrodes of 10 $\mu$m (Figure \ref{Fig_5}).

\begin{figure}[htbp]
\includegraphics[width=2.9 in]{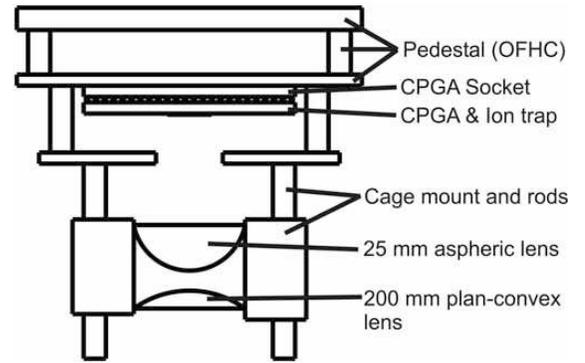}
\caption{Ion trap carrier and imaging optics ensemble. The top of the pedestal is mounted directly on to 4.2 K baseplate. (Not to scale)} \label{Fig_6}
\end{figure}

The ion traps are mounted in ceramic pin grid array (CPGA) chip carriers (Global Chip Materials, P/N PGA10047002) (Figure \ref{Fig_6}) to simplify the process of trap swapping. A CPGA is a ceramic plate with metallic pads on one side, which are used to make electrical contacts to the chip, and with a corresponding number of pins on the other side such that the CPGA can be plugged in a suitable socket or electrical board. Moreover, the CPGA's we use have over 100 usable pins which permits us to connect ion traps with a large number of electrodes.
We attach the trap to the CPGA using an epoxy resin (Varian, Torr Seal).
The trap electrodes are connected to the CPGA pads using gold wirebonds. The CPGA is inserted into a socket that is thermally anchored to the helium baseplate through an OFHC copper pedestal (Figure \ref{Fig_6}), through which electrical connections are made to the socket. To reduce the RF pickup on the DC electrodes, each DC electrode is connected to the ground through 1 nF ceramic capacitors soldered directly on the CPGA.  Since the CPGA ceramic is not a very good thermal conductor, in order to heat-sink the heat produced by the dielectric RF dissipation, we solder one end of a cooper wire mesh to the ion trap and the other end to the copper pedestal. The measured trap temperatures at various applied RF voltage amplitudes were as follows: 5.5 K at no applied voltage, 6.0 K at 200 V, 7.5 K at 400 V, and 8.5 K at 600 V.

\medskip

For applying DC bias potentials to the trap and reducing the heat transfer 36AWG phosphorus-bronze wire (Lakeshore, WSL-36-500) is used. In order to lower the heat transfer from the vacuum shield to the helium baseplate, the wires are wrapped once around the nitrogen heat shield. Before the DC wires reach the trap they go through a set of RC filters in order to reduce the electrical noise produced by the DC power supplies. The RC filters are thermally anchored to the helium baseplate with each filter having a time constant of 2 $\mu$s. As a fail safe and to check the electrical continuity of the wirebonds, every DC electrode is connected to the exterior through two wires.

The RF voltage is applied to the trap through a helical resonator.\cite{Der,Die1} The resonator is mounted on the 77 K heat shield (Figure \ref{Fig_3}) in order to reduce the capacitive load from the wire which connects the trap, and also to increase its step up. For the ion trap we used in the bath cryostat setup, the helical resonator was built such that it has a voltage step up of 20, a loaded resonant frequency of 18.45 MHz, and a quality factor of 30, as measured at room temperature. When the helical resonator is cooled at 77 K its voltage step up increases to 100, and the quality factor improves to 67. With later resonator designs we have obtained quality factors up to 300. The quality factor can be increased further if the helical resonator is thermally anchored to the 4.2 K baseplate and it is built with superconducting wire.\cite{Poi,CENT,LANC,Frunz}
A Helmholtz coil is attached to the exterior of the cryostat since we need to address the Zeeman levels of the $^{88}$Sr$^+$ ion for Doppler cooling and resolved sideband cooling. The Helmholtz coil provides 0.6 mT at an applied
current of 3 A.

\subsubsection{Vacuum system and heat load}

The ultra high vacuum in cryogenic systems can be obtained  of residual gas molecules on the cold surfaces, even in the presence of materials not compatible with room temperature UHV systems. To increase the cryosorption area, two activated charcoal getters are mounted in the cryostat, one at the top
of the nitrogen shield and another on the helium baseplate. Each getter contains about 5 g of activated charcoal enclosed
in a copper container. To cool the cryostat, we first pump it down using a turbopump while regenerating the getters by
applying 180 mW of heat through their heater coils. Once the pressure reaches $10^{-6}$ torr as measured at the turbopump inlet by a Bayard-Alpert pressure gauge, the cryostat is pre-cooled by filling both of its tanks with liquid nitrogen. When the pressure drops to $10^{-7}$ torr, the helium tank is flushed using compressed helium and a liquid helium fill is performed. The final working pressure indicated by the pressure gauge is $10^{-8}$ torr, although the pressure at the ion trap location is expected to be much lower due to the cryosorption and cryotrapping \cite{Gab} as it will be shown in Section III.

Experimentally we observe that the liquid helium hold time with the trap operating is 10 hours, and with no operation 18
hours. The hold time for nitrogen is around 12 hours regardless if the trap is operating or not. We calculated that the
heat load on the liquid helium tank with the RF turned off is 23 mW. From the liquid helium hold time, we estimate that running with typical trap parameters the heat load from the RF dielectric dissipation is 33 mW. When there are no electrical connections to the cryostat and all the windows are blocked with metallic blanks the liquid helium has about 90 hours hold time.

\subsubsection{Optical setup}

In order to run the ion trap three laser beams must be sent to the trapping location and a means for collecting the scattered photons by the trapped ions must be provided.  Laser access to the inner vacuum chamber, where the trap is mounted, is provided by three windows on the side of the
cryostat (Figure \ref{Fig_3}) and a window at the bottom of the cryostat is used for ion imaging. Each shield of the cryostat has a set of windows.  There is an outer set mounted on the room temperature vacuum shield
which holds the vacuum in the cryostat while allowing optical access, and an inner set on the 77K shield whose purpose is
to absorb 300 K radiation and sink it into the nitrogen, increasing the helium hold time. For the outer set, we use 50 mm BK7 windows (Melles Griot, 02WBK226), and for the inner set 1 inch BK7 windows (Thorlabs, WG11050). All the access windows used in the bath cryostat system have no anti-reflection coatings.

For imaging the ions an aspheric lens is placed above the trap to collimate the fluorescence from the ions. The aspheric
lens was chosen such that it had a good light collection and it can be positioned far enough from the trap so that the
electrostatic charges on it do not affect the ions. We opted for an aspheric lens made by Edmund Optics (NT47-730) with
an EFL of 25 mm and f/1, which collects 5.3\% of the light emitted by the trapped ions. The aspheric lens and a 200 mm plano-convex lens which focus the image outside the cryostat, are placed in a 1 inch cage mount from Thorlabs. The entire ensemble can slide on four cage rods as shown in Figure \ref{Fig_6}.  Although there is a mismatch in the thermal expansion coefficients between the lenses and cage mount, the lenses have survived hundreds of cool down cycles. The final image is focused onto EMCCD camera (Princeton Instruments, IMAX/PHOTONMAX 512) by two 100 mm plano-convex lenses placed outside the cryostat. The imaging system has 4$\times$ magnification and an overall measured resolution of 6 $\mu$m.

\begin{figure}[htbp]
\includegraphics[width=2.2 in]{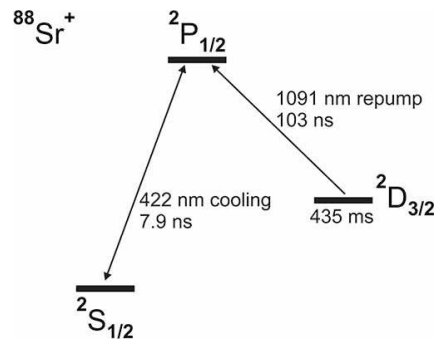}
\caption{Partial energy levels of $^{88}$Sr$^+$. The S$_{1/2} \leftrightarrow $P$_{1/2}$ transition is used for Doppler cooling \cite{Man}. The population decayed to the metastable D$_{3/2}$ level is repumped back to P$_{1/2}$ level by the 1091 nm laser.} \label{Fig_7}
\end{figure}

The energy levels of $^{88}$Sr$^+$ and the transitions used to conduct the experiments reported in this paper are shown in Figure \ref{Fig_7}. The S$_{1/2} \leftrightarrow $P$_{1/2}$ transition is used for Doppler cooling the trapped ions and is driven by a 422 nm laser with 20 $\mu$W power focused to a 33 $\mu$m spot size. The P$_{1/2}$ state has a probability of
1 in 13 to decay to the metastable D$_{3/2}$ level (435 ms lifetime). To avoid the depopulation of the Doppler cooling transition, we use a
1091 nm repumping laser with 50 $\mu$W power focused to a 100 $\mu$m spot size. The light necessary to drive these
transitions is
produced from two external cavity diode lasers with optical feedback for frequency stabilization. The details of the lasers
used in this experiment are described elsewhere.\cite{Lab}

This liquid helium bath cryostat system was used to cool a single trapped ion to the motional ground state with a 95$\%$ fidelity.\cite{Jarek}

\subsection{Closed cycle cryostat}
\label{subsec:ccycle}

\begin{figure}[htbp]
\includegraphics[width=2.9 in]{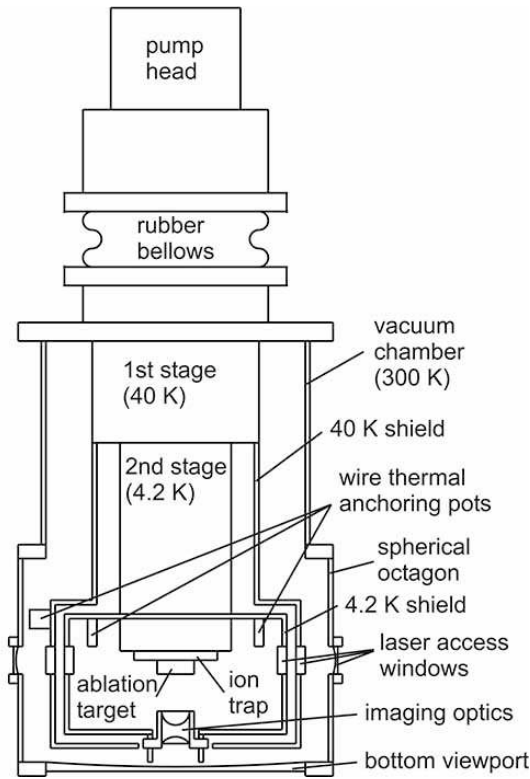}
\caption{Schematic cross section through the closed cycle cryostat. The imaging optics anchored onto 4 K radiation shield is exposed directly to the 300 K radiation from the bottom viewport. (Not to scale)} \label{Fig_8}
\end{figure}

The closed cycle cryostat system is constructed around a low vibration cryostat produced by Advanced Research Systems (Model GMX-20B). The cryocooler head of the cryostat is mechanically decoupled from sample tip and vacuum chamber around it. The cryostat interface has two cooling stages, where the minimum temperature of the first stage is 40 K, and the minimum temperature of the second stage is 4.2 K. The cryostat interface comes with an 8 inch conflat (CF) flange and the UHV chamber is built by attaching a full CF nipple, a spherical octagon (Kimball Physics, Model MCF800-SO2000800-A) and an 8 inch glass viewport. The experiment chamber is made from an OFHC cooper tube attached to the 4.2 K cold tip. The cold tip and experiment chamber are surrounded by a custom made radiation shield attached to the 40 K stage of the cryostat (Figure \ref{Fig_8}). Only the room temperature outer chamber is vacuum sealed, with the inner chambers non-hermetically closed. The temperatures of the 4.2 K and 40 K stages are monitored with two silicon diodes (LakeShore, DT-670A-SD).

One concern due to the principle of operation of the closed cycle cryostats, is the vibration transmitted to the ion trap during experiments. Since the cryocooler head is mechanically decoupled from the cryostat interface, it must be supported such that the vibration transfer to the cryostat interface is minimized. In our setup we chose to place the cryostat on an optical table with the cryocooler head supported independently by a holder anchored to the ceiling.

\begin{figure}[htbp]
\includegraphics[width=3.2 in]{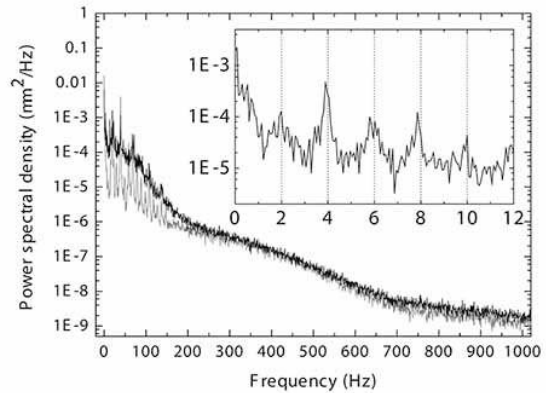}
\caption{Vibration power spectrum of the closed cycle cryostat. The vibration power spectrum was measured with the cryostat compressor turned off (gray curve) and on (black curve). The inset shows the lower part of the spectrum where the fundamental and higher harmonics of the cryocooler head vibration can be distinguished. The spectrum has 2.5 kHz bandwidth and 0.17 Hz resolution.} \label{Fig_10}
\end{figure}

The vibration amplitudes of the cold tip were measured with a Michelson interferometer with the light provided by the 422 nm laser. During the operation of the cryocooler compressor we found that the maximum displacement of the cold tip was below 106 nm (one quarter of an interference fringe). A typical vibration power spectrum of the cold tip is shown in Figure \ref{Fig_10}. The cryocooler head has a pumping repetition rate of 2 Hz, which can be observed in Figure \ref{Fig_10} inset.

\medskip

\begin{figure}[htbp]
\includegraphics[width=2.9 in]{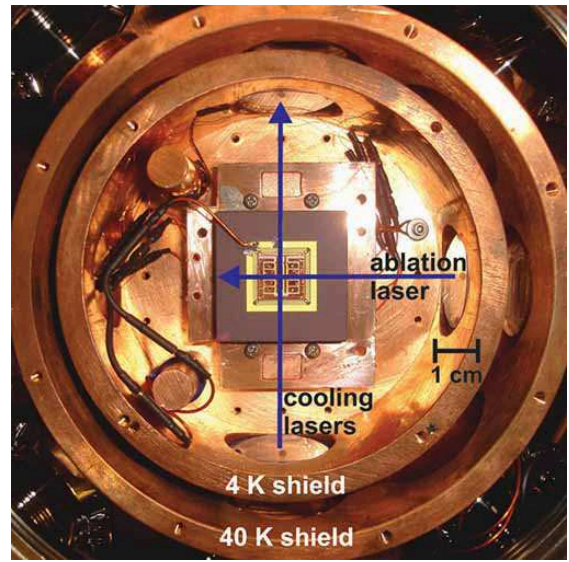}
\caption{View of the closed cycle cryostat working chamber. The ablation target is not installed; the rest of visible components are described in Figure \ref{Fig_8}. Note: The ion trap employed in this setup is different from the one used in the bath cryostat system (Figure \ref{Fig_5}). For a detailed description of this ion trap consult reference \cite{David}.} \label{Fig_9}
\end{figure}

\subsubsection{Ion trap and electrical components}

The ion trap used in this setup was built by patterning the copper electrodes on a printed circuit board.\cite{David,Pea} The ion trap and its CPGA are installed on a pedestal which is identical with the one used in the bath cryostat setup. The pedestal is thermally anchored to the 4.2 K cold tip. We found that the temperatures of the ion trap as function of the applied RF voltage amplitude were: 7.2 K at no applied voltage, 8.8 K at 500 V, and 9.4 K at 650 V.

DC bias potentials are applied to the trap through the same phosphorus-bronze wires used in the bath cryostat.  The RF potential is delivered through a 20 AWG copper wire, since for this setup the helical resonator is mounted outside the cryostat. To heatsink the heat conducted by the wires from exterior, the wires are wrapped around copper posts mounted on the 40 K cooling stage heatshield and on the 4.2 K baseplate.

\subsubsection{Vacuum system and heat load}

To attain the UHV environment the cryostat is first pumped with a turbomolecular pump to $5 \times 10^{-7}$ torr measured by an inverted magnetron pressure gauge attached at the pump inlet. In the second stage the cryostat vacuum chamber is isolated from the turbomolecular pump and the cool down process is started, while the pumping is continued by a small ion pump attached directly to the spherical octagon. The cooling of the cryostat takes 4 hours, with the cold stage reaching a final temperature of 5.2 K and the pressure at the ion pump getting to $3 \times 10^{-9}$ torr. The computed heat load from radiation and conduction to the cold tip is about 1 W, which agrees well with cryostat heat load map supplied by the manufacturer. The highest heat load (0.4 W) is from the 300 K radiation absorbed by the exposed lenses assembly. In the future we plan to decouple the two lenses and mount the 75 mm lens on the first stage bottom plate. For comparison the thermally unloaded cold tip reaches 4.2 K after 2 hours of cooling.

\subsubsection{Optical setup}

To allow for laser access to the trap, a set of three 2.75 inch viewports are attached on the side of the spherical octagon. The access through the 40 K and 4.2 K heat shields is provided by six 1 inch BK7 windows (Thorlabs, WG11050), with three windows mounted on each shield. The imaging of the ions is done through the bottom 8 inch viewport. The light scattered by the trapped ions is collected and collimated by an aspheric lens (Edmund Optics,  NT49-100, 22.50 mm EFL, f/1.50). The collimated light is focused outside the cryostat by a 75 mm  plano-convex lens. Both lenses are mounted in 1 inch Thorlabs cage mount which can slide on four posts attached to the bottom plate of the cryostat second stage chamber (Figure \ref{Fig_8}). Two lenses, a 75 mm and a 150 mm achromatic doublet are installed outside the cryostat, to focus the final image on a CCD camera (SBIG, ST-402ME). The entire imaging system has a 7.5$\times$ magnification and provides a theoretical resolution of 1.2 $\mu$m. For this setup we use a laser system similar to the one used for the bath cryostat setup.

\section{Performance Evaluation}
\label{sec:eval}

\subsection{Ion loading efficiency}
In room temperature systems there are several methods for ion loading. In our systems we successfully tested the following methods: the neutral atoms produced by a resistively heated oven are ionized by impact with electrons emitted by an e-gun or through laser photoionization, and the ions are directly produced by laser ablation of a solid target or indirectly by photoionization of the neutrals inside the ablation plume. For our experiments, we chose the laser ablation loading method as it has some advantages compared with the other loading methods. During the ion loading from an oven, the heat production is of the order of hundreds of Joules, but in comparison a single ablation pulse of 1.5 mJ can load our trap. If the ion trap were operated in a cryostat which could reach temperatures in the milikelvin range, the system would likely not be able to handle the dissipated heat from an oven. Another benefit of laser ablation is that it can produce ions from materials with very high melting temperature. Surface-electrode traps tend to have a lower trap depth than a linear RF ion trap of comparable dimensions, and thus provide a stringent test for the laser ablation ion loading method.\cite{David}

The mechanism of ion trapping from the ablation plume was described by Hashimoto {\it et al.}.\cite{Has} The fast electrons from the ablation plume arrive first to the trap and cancel the RF voltage applied to the trap, thus allowing the slower moving ions to enter the trapping region even if their kinetic energy is lower than the trap well depth. To produce the ablation plume, we use for both cryostat setups a Q-switched Nd:YAG laser (Continuum Electro-Optics, Minilite II), which produces a maximum energy in its third harmonic (355 nm) of 8 mJ per pulse. The pulse duration is around 5 ns. The beam is focused onto the ablation target to spot size of 0.5 mm. Because metallic $^{88}$Sr easily oxidizes in
open atmosphere a single crystal of SrTiO$_3$ (MTI Crystal, STOa100505S1) placed 15 mm from the center of the trap
(Figure \ref{Fig_3}) is used as ablation target. The YAG beam is steered by a high-energy mirror (CVI Laser, Y3-1025-45-P). The energy loss in optics is around 1 mJ, thus the maximum energy delivered to the target is 7 mJ as measured with a pyroelectric meter.

\begin{figure}[htbp]
\includegraphics[width=2.9 in]{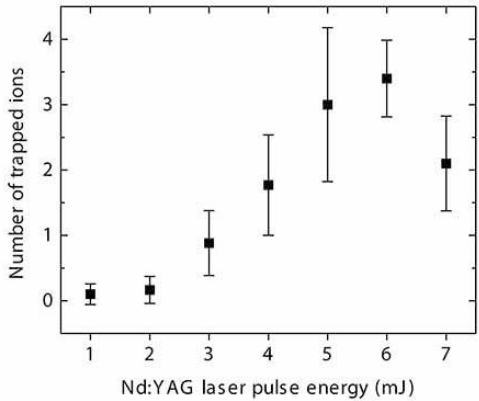}
\caption{Ion loading efficiency versus ablation laser pulse energy in the bath cryostat system.} \label{Fig_11}
\end{figure}

In order to study the ion loading efficiency from the ablation plume in the bath cryostat setup, the energy of the Nd:YAG pulse on the target was varied from 1 mJ to 7 mJ (1 MW/mm$^2$ to 7 MW/mm$^2$). The trap was operated at an applied RF voltage of 200 V at a frequency of 18.45 MHz, with a trap depth of 0.34 eV, and with the ions located at 200 $\mu$m above the trap. The endcap DC electrodes were maintained at 7.5 V and the middle DC electrodes were maintained at -5 V (Figure \ref{Fig_5}). The secular frequencies of the trap were as following: 550 kHz, 1.52 MHz, and 1.87 MHz. During the experiment, the Doppler cooling laser was detuned 30 MHz below the transition frequency. The maximum efficiency in ion loading was obtained at 6 mJ laser pulse energy as shown in Figure \ref{Fig_11}.

\subsection{Ion lifetime}
One of the primary reasons to choose a cryogenic systems is that the vapor pressure of most gasses is negligible allowing them to reach very high vacuum in a short time. Most of the residual background gas in the bath cryostat comes from the 77 K heat shield and in the closed cycle cryostat from the 4 K shield.  The residual gas species with an appreciable vapor pressure at 77 K are He, O$_2$, N$_2$, and H$_2$ and at 4.2 K only He and H$_2$.\cite{Hae}

The ion lifetime is typically limited by collisions, charge exchange, or chemical reactions with residual background gas.\cite{Rot,Win,Sug,Ray} The elastic collisions cannot transfer enough kinetic energy to the Doppler cooled trapped ion to eject it out of the trap, since the residual molecules have a kinetic energy no higher than 77 K and they are lighter then the $^{88}$Sr$^+$ ion, which is trapped at 3500 K (0.3 eV) depth in the bath cryostat or 4700 K (0.4 eV) in the closed cycle cryostat. The ionization energies of the residual molecules and atoms (He 24.58 eV, O$_2$ 12.30 eV, N$_2$ 15.60 eV, H$_2$ 15.20 eV \cite{Chem,CRC}) from the background gas are higher then the ionization energy of the $^{88}$Sr atom (5.69 eV). Even if it is assumed that the $^{88}$Sr$^{+}$ ion is in the excited P$_{1/2}$ state and the energy of the 422 nm photon (2.97 eV) is transferred towards the ionization of the colliding atom or molecule the charge exchange reactions cannot occur. Chemical reactions between the ion and the background gas remain as the only possibility through which an Doppler cooled ion held in a cryogenic trap can be lost. Among all the possible chemical reactions between the $^{88}$Sr$^{+}$ ion and the residual gas,\cite{Rot,Win,Sug,Ray,Chem,CRC} in which the kinetic energy of the final $^{88}$Sr compound is comparable with the trapping potential depth, is:

\vspace{10pt}

\noindent $^{88}$Sr$^{+*} + $O$_2 \rightarrow ^{88}$SrO$^+ + $O$ + 1.80$ eV,

\vspace{10pt}

\noindent where $^{88}$Sr$^{+*}$ is the ion in the P$_{1/2}$ state. In the above reaction the SrO$^+$ has a kinetic energy of 0.3 eV which is enough to escape from the trap operated in the bath cryostat system.

\medskip

\begin{figure}[htbp]
\includegraphics[width=2.9 in]{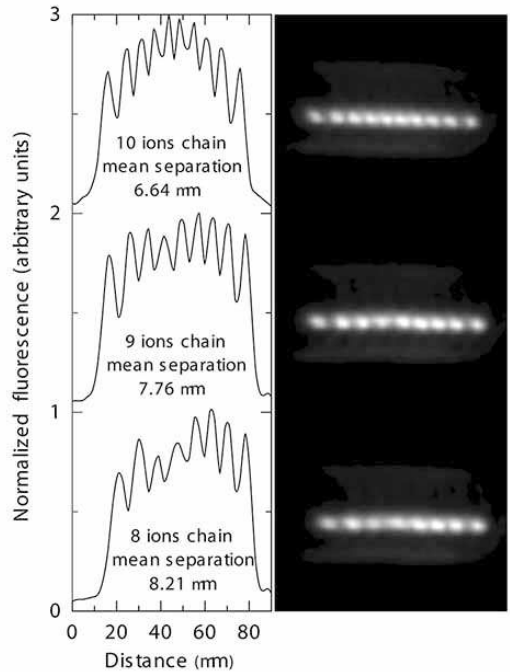}
\caption{The ion chain used to determine the lifetime of a single ion in the bath cryostat system. The right panel shows the initial chain of 10 ions and the subsequent two chains where one ion, and two ions respectively, were lost.} \label{Fig_12}
\end{figure}

To determine the lifetime of a Doppler cooled single ion trapped in the liquid helium bath cryostat setup, we observed the rate at which ions were lost from an initial chain of 10 ions. Experimentally, we found a lifetime of 680 minutes for a single ion. During the experiment no ion became a non-resonant ion due to chemical reactions with the residual background gas (Figure \ref{Fig_12}). The Doppler cooling laser (422 nm) and the repumping laser (1091 nm) intensities were chosen such that the $^{88}$Sr$^{+}$ ion spent approximately a third of the time in the P$_{1/2}$ excited state. Without Doppler cooling we measured a single ion lifetime of 400 s.

Knowing the lifetime of a single atom trapped in the bath cryostat system and the collision rate between the trapped ions and residual molecules, an estimate of the O$_2$ partial pressure at the trap location can be made. The collision rate can be found from the classical Langevin theory of ion-molecule collisions:\cite{Win,Ray}

\[\Gamma=\frac {ne}{2\epsilon_0} \sqrt{\frac {\alpha}{\mu}},\]

\noindent where $n$ is the density of the background gas molecules, $\alpha$ is the polarizability of the molecules, and
$\mu$ is the reduced mass of ion-molecule system. By assuming that the background gas has a temperature of 77 K and
neglecting the kinetic energy of the trapped ions, we estimate an upper bound on O$_2$ partial pressure of $2\times10^{-12}$ torr.

\medskip

\begin{figure}[htbp]
\includegraphics[width=2.9 in]{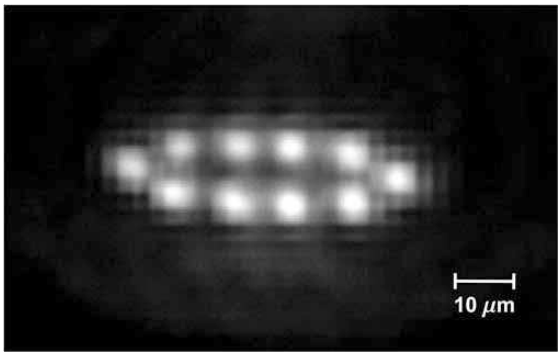}
\caption{A typical small ion crystal trapped in the closed cycled cryostat system. The ion crystal lies 750 $\mu$m above the ion trap at a trapping potential depth of 0.40 eV.} \label{Fig_13}
\end{figure}

If residual gas is not chemically reactive with the ion then, they can still interact via elastic collisions. Collisions can heat or cool the ions depending on the mass and temperature of the buffer gas.\cite{Mori1,Mori2,Mori3} Injecting He buffer gas provides a means of cooling the ions to cryogenic temperatures that is insensitive to the ion state and does not require laser cooling. We characterized the ion lifetime without laser cooling in the presence of He buffer gas at a partial pressure of $7\times10^{-6}$ torr and at a temperature of 7 K using the closed cycle cryostat. After 12 hours we found there was no discernible drop in the ion cloud fluorescence intensity compared with its initial intensity.

To characterize the ion lifetime in the absence of He buffer gas and with laser cooling in the closed cycle cryostat we employed the same method as for the bath cryostat. We observed an ion chain of four ions for a period of 12 hours in which none of the ions got lost or became non-fluorescent. This sets a lower bound for a single ion lifetime to 2500 minutes, which is in agreement with an unlimited ion lifetime obtained from a theoretical estimate similar to the one made for the bath cryostat system.

\section{Conclusion}
 We have shown that the cryogenic ion traps offer many advantages and we successfully managed some problems related to the operation of ion traps in cryogenic environments. The ion traps can be swapped very quickly and UHV pressures can be reached in less then 12 hours, which decreases the time necessary to implement new trap designs. Also new materials which are not room temperature UHV compatible can be used in the design of these cryogenic systems. The pressures obtained inside these cryogenic systems permit for the Doppler cooled ions to have a trapped lifetime of 680 minutes for the bath cryostat setup and over 2500 minutes for the closed cycled cryostat setup. To load the ion traps we employed an ion source based on laser ablation. By this method we can load surface-electrode ion traps with depths as low as 0.3 eV. In addition, the low temperature of these cryogenic systems offers the possibility of interacting the ions with superconducting systems.



\end{document}